\documentclass[floatfix,10pt,onecolumn,showpacs,amsmath,amssymb]{revtex4}

\usepackage{epsf}
\usepackage{graphicx}  % Include figure files
\usepackage{dcolumn}   % Align table columns on decimal point
\usepackage{bm}        % bold math

\newcommand{\be}{\begin{equation}}
\newcommand{\en}{\end{equation}}
 \newcommand{\bea}{\begin{eqnarray}}
 \newcommand{\ena}{\end{eqnarray}}
  \newcommand{\sch}{Schwarzschild}

\begin{document}

\title{Twisted spacetime in Einstein gravity}
\author{Hongsheng Zhang$^{1,2~}$\footnote{Electronic address: hongsheng@shnu.edu.cn} }
\affiliation{ $^1$Center for
Astrophysics, Shanghai Normal University, 100 Guilin Road,
Shanghai 200234, China\\
$^2$State Key Laboratory of Theoretical Physics, Institute of Theoretical Physics, Chinese Academy of Sciences, Beijing, 100190, China
}

%\date{ \today}

\begin{abstract}
  We find a vacuum stationary twisted solution in four-dimensional Einstein gravity. Its frame dragging angular velocities are antisymmetric with respect to the equatorial plane. It possesses a symmetry of joint inversion of time and parity with respect to the equatorial plane. Its Arnowitt-Deser-Misner (ADM) mass and angular momentum are zero. It is curved but regular all over the manifold. Its Komar mass and Komar angular momentum are also zero. Its infinite red-shift surface coincides with its event horizon, since the event horizon does not rotate. Furthermore we extend this solution to the massive case, and find some similar properties. This solution is a stationary axisymmetric solution, but not Kerr. It explicitly proves that pure Einstein gravity permits different rotational mode other than Kerr.  Our results demonstrate that the Einstein theory may have much more rich structures than what we ever imagine.
\end{abstract}

\pacs{04.20.-q, 04.70.-s}
\keywords{black hole; Einstein gravity}

%\preprint{arXiv: }
 \maketitle

\section{Introduction}
Einstein's general relativity \cite{ein} is the standard modern gravity theory. Exact solutions take  pivotal status in the theory. The well known three classical tests of general relativity depend on the \sch solution \cite{schw}, which play decisive roles in the early developments of the general relativity. Many exact solutions for Einstein equation have been found \cite{exa}.
 Among these exact solutions, two of the most significant solutions  are \sch~and Kerr \cite{kerr}. The \sch~solution describes the spacetime around a spherically symmetric star. The Kerr solution describes the spacetime around a rotating star. Actually, the \sch~and Kerr solutions can describe the gravitational fields in vast ranges from millimetre to the clusters of galaxies (the Newton's law as a weak field limit).

It is generally believed that the \sch~ solution is the unique static vacuum solution, and  the Kerr solution is the unique rotating vacuum solution on an asymptotic manifold in Einstein gravity after the proof of the uniqueness theorems, for reviews see \cite{uni}. Almost all the celestial objects are rotating, quickly or slowly. So the Kerr solution is extremely important in the studies of astrophysics.  Unfortunately we cannot obtain a completely satisfactory
    interior solution matching to Kerr up to now, though great efforts have been evolved in this topic for more than 50 years. On the contrary, we obtain several negative results. For examples, a perfect fluid cannot be the source of Kerr \cite{her}, and the analytic approximate solution that describes the slowly rotating astrophysical object does not lead to Kerr \cite{bos}. The other problem of the Kerr solution is that it is unstable against linear perturbations in the interior region \cite{teu}. That means the result of collapsing of a rotating star may not be a Kerr black hole, even if the exterior region of the progenitor can be described by Kerr solution.  So, theoretically the question is: Are there any rotating modes different from Kerr, more or less like the wave functions of the hydrogen atom with the same angular momentum?

     At the first sight the answer is no, since the Carter uniqueness theorem forbids them \cite{car}. Exactly, the Carter uniqueness theorem requires: a. the manifold is axially symmetric and stationary;  b. the manifold is asymptotically flat, and the total mass and total angular momentum measured at the infinity, i.e., the  ADM mass and angular momentum \cite{adm}, are $M,~J$ respectively; c. the manifold is regular everywhere at the exterior region of the horizon (including the horizon). Under these conditions, the spacetime must be Kerr. This is a quite exciting but harsh theorem. It almost determines the metric around any celestial object (almost all  celestial bodies are rotating), including planet, stars, galaxies, and cluster of galaxies. It is just Kerr. However, as we have mentioned, Kerr has some problems. Interestingly, by a careful analysis of the Israel uniqueness theorem \cite{isr} and the Carter uniqueness theorem, we find that the Einstein equation  permits more vacuum asymptotically flat rotating solution other than Kerr.

     We explain the general idea for our approach on a rotating spacetime to evade the uniqueness theorems. We consider a rotating spacetime with $J=0$. Since it is rotating the Israel uniqueness theorem say nothing about this case. At the same time its total angular momentum is zero, thus the Carter uniqueness neither requires it to be Kerr. Our key idea is that ``rotating" does not contradict with $J=0$, which is beyond the traditional lore. For example, we can consider the case that the space rotates at different directions in different spherical shells. The angular momentums in different shells are arranged to be exactly counteracted,  and thus the total angular momentum vanishes. Without breaking the axial symmetry, we require the directions of rotation of the shells to be up or down. We find a twisted solution which is a little more complicated than this heuristic example.

The metric read,
\be
ds^2=-\frac{r^2-a^2}{r^2+a^2}dt^2+\frac{4a(r^2-a^2)\cos\theta}{r^2+a^2}dtd\phi+\frac{r^2+a^2}{r^2-a^2}dr^2+(r^2+a^2)d\theta^2+
 \frac{4a^4-4a^2r^2-(3a^4-6a^2r^2-r^4)\sin^2\theta}{r^2+a^2}d\phi^2,
 \label{met1}
 \en
where $a$ is a constant, and ($t,~r,~\theta,~\phi$) are spherical coordinates, which come back to the standard spherical coordinates in Minkowski spacetime when $a=0$. The Ricci tensor for this metric vanishes,
\be
R_{ab}=0.
\en
Thus it is a vacuum solution of the Einstein equation. It is a curved space and regular all over the manifold, since
 the only non-zero scalar polynomial of curvatures (the Kretschmann scalar) reads,
 \be
 R^{abcd}R_{abcd}=\frac{48(a^8-15a^6r^2+15a^4r^4-a^2r^6)}{(r^2+a^2)^6}.
 \label{kre}
 \en
 It is easy to check that the spacetime (\ref{met1}) possesses a discrete symmetry of joint transformations of time inversion ($t\to~ -t$) and reflection with respect to the equatorial plane ($\theta\to~\pi-\theta$).

The infinite red-shift surface dwells at
\be
g_{00}=0.
\en
 And the event horizon $f=$constant satisfies,
 \be
 g^{{\mu}{\nu}}\frac{\partial f}{\partial x^\mu}\frac{\partial f}{\partial x^\nu}=0.
 \en
 It is easy to obtain that the infinite red-shift surface coincides with the event horizon at $r=a$.  We shall discuss the the reason for this coincidence later. It is a black hole, since it has an event horizon. However, it is very different from the ordinary holes, which have spacetime singularities. From (\ref{kre}) one sees that the black hole (\ref{met1}) is regular everywhere in the whole spacetime.

 The frame dragging angular velocity reads,
 \be
 \Omega=-\frac{g_{03}}{g_{33}}=\frac{2 a \left(a^2-r^2\right) \cos\theta}{4a^4-4a^2r^2-(3a^4-6a^2r^2-r^4)\sin^2\theta}.
 \en
 Based on the detailed discussions on the rotating metrics, one finds that the frame dragging velocity can be treated as the angular velocity of the spacetime itself \cite{rot}. A simple explanation is that a co-rotating observer,
  \be
  \frac{\partial}{\partial t}+\Omega\frac{\partial}{\partial \phi},
  \en
  will sense a time-orthogonal spacetime like a static one (without space-time cross term).
  At the horizon $r=a$, the frame dragging velocity vanishes. In this sense, the horizon is static. Thus the infinite red-shift surface coincides with the event horizon. That is different from the case of Kerr, for which the event horizon is rotating, so that the infinite red-shift surface is separated from the event horizon. The other important property of the frame dragging angular velocity is that it vanishes at the equatorial plane $\theta=\pi/2$, and is antisymmetric with respect to the equatorial plane, i.e., $\Omega(\theta)=\Omega(\pi-\theta)$.

  Fixing $\theta$, for the large $r$ approximation we have,
 \be
 \Omega\sim \frac{1}{r^2}.
 \en
 Hence, $\Omega$ has at least one maximum for $r\in [a, \infty)$. In this interval only one point satisfies,
 \be
 \frac{d\Omega}{dr}=0,~~ {\rm and}~~\frac{d^2\Omega}{dr^2}<0.
 \en
 At this point,
\be
 r=\sqrt{3}a,~~ {\rm and}~~\Omega=\frac{\cos\theta}{2a(1-3\sin^2\theta)}.
 \en
From the previous discussions, we obtain the picture of this spacetime. It is a twisted spacetime. The spacetime rotates in opposite directions  above and below the equatorial plane. The equatorial plane itself does not rotate. A  sketch of this spacetime is shown in the fig. 1.   If one inserts an elastic bar with finite thickness along $\theta=0$, it will be twisted to be something like a screw steel bar.

\begin{figure}
\begin{center}
\includegraphics[scale=0.8]{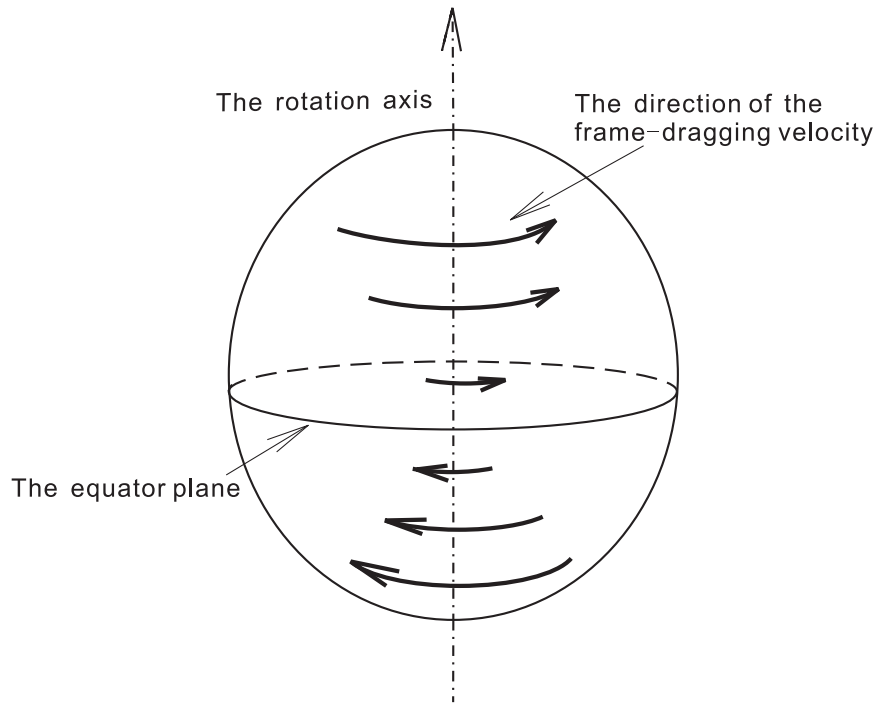}
\caption{The illustration of a spherical surface (not the horizon) in the twisted spacetime (\ref{met1}). The frame-dragging angular velocities are antisymmetric with respect to the equatorial plane. The equatorial plane is static. }
\end{center}
\label{twbh}
\end{figure}

Next we study the mass and angular momentum of this spacetime. First we show that the spacetime (\ref{met1}) is asymptotically flat. For the sake of canceling the frame dragging effects, we introduce the co-rotating coordinates,
\be
\bar{t}=t,~\bar{r}=r,~\bar{\theta}=\theta,~{\rm and}~\bar{\phi}=\phi-\Omega t.
\label{dra}
\en
With this new coordinate system, the metric (\ref{met1}) reads,
\be
ds^2=(g_{00}-g_{03}^2/g_{33})d\bar{t}^2+g_{11}d\bar{r}^2+g_{22}d\bar{\theta}^2+g_{33}(d\bar{\phi}+\bar{t}d\Omega)^2,
\en
where $g_{\mu\nu}$ represent the exponents of metric (\ref{met1}). Here they are treated as the functions of ($\bar{t},~\bar{r},~\bar{\theta},~\bar{\phi}$). One can show that ($\bar{t},~\bar{r},~\bar{\theta},~\bar{\phi}$) are (quasi-)spherical coordinates  in the ADM formalism, which satisfies
  \be
   g_{\mu\nu}=\eta_{\mu\nu}+{\cal O}(\frac{1}{\bar{r}}),
   \en
  where $\eta_{\mu\nu}$ are the components of Minkowski metric in spherical coordinates. According to the standard formulae of ADM mass and ADM angular momentum, we obtain
  \be
  M_{\rm ADM}=0,
  \label{admm}
  \en
  and
   \be
  J_{\rm ADM}=0.
  \en
 The ADM formulism can only obtain the total mass and angular momentum. It cannot say anything about the spatial distributions of the mass and angular momentum.
 The distribution of mass and angular momentum of gravity field is a very intricate problem. A local mass seems necessary in the cases, for example, the propagation of gravitational waves from the source to the local observer. However, an energy-stress density is prohibited by the equivalence principle. Thus
 we turn to the quasi-local forms as an inevitable  concession. After decades' studies we have several different definitions quasi-local gravitational mass and angular momentum. Usually, they are not equal to each other in the same 2-surface. Some of them are significant not only in the studies of mass of gravity, but also in other topics including thermodynamics, exact solutions etc \cite{self}.  Here we do not check them one by one for the metric (\ref{met1}).
   As an example we study one of the earliest form, the Komar integral, which is applicable in the stationary spacetime \cite{kom}. The Komar integral defines the gravitational mass by the imprints of gravitational effects on a 2-surface in a stationary spacetime. Originally, it is presented as the total mass for a space-like
   world sheet in an infinite 2-surface. In principle, it is also can be used to define a quasi-local mass in a finite 2-surface,
   \be
   M_{\rm Komar}=-\frac{1}{8\pi}\int *d\xi,
   \label{komar}
   \en
   where a star denotes the Hodge dual operator, and $\xi$ is the lower index form of the time like Killing vector,
   \be
   \xi_a=g_{ab}\left(\frac{\partial}{\partial t}\right)^b.
   \en
   For the metric (\ref{met1}), the Komar integral (\ref{komar}) in 2-surface of radius $r$ presents,
   \be
   M_{\rm Komar}=\frac{2a^2r}{a^2+r^2}.
   \en
   The Komar mass vanishes when $r\to \infty$. This result is consistent with the ADM mass (\ref{admm}). The physical interpretation is that the
   kinetic energy (because of rotation) exactly counteracts the potential energy. For a finite 2-surface, the Komar mass always larger than zero. This means that
   the kinetic energy always larger than the potential energy. The Komar angular momentum reads,
   \be
    J_{\rm Komar}=\frac{1}{16\pi}\int *d\xi_{\phi},
   \label{komara}
   \en
   where
   \be
   ({\xi_{\phi}})_a=g_{ab}\left(\frac{\partial}{\partial \phi}\right)^b.
   \en
   In a 2-surface of radius $r$, the Komar angular momentum reads,
   \be
    J_{\rm Komar}=\frac{3a^3r-ar^3}{2(a^2+r^2)}\int \sin\theta\cos\theta d\theta=0.
    \en
   This result is also consistent with the ADM angular momentum. Further, it presents a more fine result, which is independent of $r$. The Physical interpretation is that the angular momentum above the equatorial plane exactly counteracts the angular momentum
   below the plane. Of course, it embodies the property of antisymmetry of the frame-dragging angular velocity with respect to the equatorial plane. So, we call this solution rotating space without (total) mass and (total) angular momentum. There is only one parameter $a$ in this solution, which denotes the
   angular velocity of the spacetime. In the following text, we shall introduce one more parameter into the metric (\ref{met1}), which can be interpreted as
   the total mass of a spacelike world sheet.

   Then we make a preliminary discussion about the motion of a test particle in the spacetime (\ref{met1}). First we consider a particle moving on the equator
   plane $\theta=\pi/2$. For a time-like geodesics, we obtain at large $r$ approximation,
   \be
   \frac{d^2\rho}{d\phi^2}+\rho=\frac{2a^2\rho}{L^2}+4a^2\rho^3.
   \en
   Here $\rho=1/r$, and $L$ is the angular momentum of the test particle. Note that $L$ is different from the angular momentum of the spacetime $J$. Comparing to the equation of motion in the \sch~spacetime, we find that the effective \sch~mass is $2a\rho$, and the leading correction term becomes $1/r^3$ rather than $1/r^2$.

   For more realistic studies, we extend the solution (\ref{met1}) to a solution with one more parameter $M$,
   \bea
ds^2=-\frac{r^2-a^2-2Mr}{r^2+a^2}dt^2+\frac{4a(r^2-a^2-2Mr)\cos\theta}{r^2+a^2}dtd\phi+\frac{r^2+a^2}{r^2-a^2-2Mr}dr^2+(r^2+a^2)d\theta^2 \nonumber\\+
 \frac{4a^4+8a^2Mr-4a^2r^2-(3a^4+8a^2Mr-6a^2r^2-r^4)\sin^2\theta}{r^2+a^2}d\phi^2.
 \label{met2}
 \ena
    The metric (\ref{met2}) is a solution of vacuum Einstein equation. For it the Ricci tensor reads,
      \be
      R_{ab}=0.
      \en
            It is easy to see that the metric (\ref{met2}) comes back to metric (\ref{met1}) when $M=0$. Also one can check that (\ref{met2}) becomes \sch~ metric when $a=0$. The infinite red-shift surface and event horizon coincide at,
            \be
            r=M+\sqrt{M^2+a^2}.
            \en
            The metric (\ref{met2}) is still regular all over the spacetime, since the Kretschmann scalar reads,
            \bea
       R^{abcd}R_{abcd}=~~~~~~~~~~~~~~~~~~~~~~~~~~~~~~~~~~~~~~~~~~~~~~~~~~~~~~~~~~~~~~~~~~~~~~~~~~~~~~~~~~~~~~~~~~~~~~~~~~~~~~~~~~~~~~~~~~~~~~~~~~~~~~~\nonumber\\
             \frac{48(a^8-a^6M^2+12a^6Mr+15a^4M^2r^2-40a^4Mr^3-15a^2M^2r^4+12a^2Mr^5-15a^6r^2+15a^4r^4-a^2r^6+M^2r^6)}{(r^2+a^2)^6}.
           \label{kre2}
            \ena
             The properties of the frame-dragging velocity are also similar to the massless case. In the dragged frame (\ref{dra}), one can demonstrate the asymptotic flatness of the spacetime (\ref{met2}). Then the ADM mass and angular momentum read,
      \be
      M_{\rm ADM}=M,
      \en
       and
       \be
       J_{\rm ADM}=0,
       \en
       respectively.
      The Komar mass and angular momentum in a 2-surface of radius $r$ read,
        \be
        M_{\rm Komar}=\frac{a^2(2r-M)+Mr^2}{r^2+a^2},
        \en
       and
       \be
       J_{\rm Komar}=\frac{a^3(3r-M)+r^2a(3M-r)}{2(a^2+r^2)}\int \sin\theta\cos\theta d\theta=0,
       \en
       respectively. We see that $M$ is the mass parameter, which reduces to the \sch mass when $a=0$, and the total angular momentum is zero. The physical interpretations of these results follow the previous case. Mimicking the previous massless case, we obtain the equation of geodesics for a test particle moving on the equatorial plane at large $r$ approximation,
       \be
       \frac{d^2\rho}{d\phi^2}+\rho-3M\rho^2-\frac{M}{L^2}=\frac{2a^2\rho}{L^2}+4a^2\rho^3.
       \label{mov2}
       \en
       In this form, we recover all the terms for the \sch~ solution (the terms at the left hand side of the above equation). The correction terms
       relative to the \sch~spacetime are casted to the right hand side of the (\ref{mov2}).

       In summary, we obtain a vacuum stationary asymptotically flat axisymmetric twisted spacetime in Einstein gravity and study some preliminary properties of it. It describes a rotating spacetime with total angular momentum $J=0$. The directions of the angular velocities of the spacetime are antisymmetric
       with respect to the equatorial plane. So its concrete image is a twisted spacetime. There is no true singularity in this spacetime. Its ADM mass and angular momentum, and total Komar mass and angular momentum are zero. We study the motion of a test particle on the equatorial plane and find the equation of motion. Finally, we present a massive version of this twisted solution.  Usually, the Carter uniqueness theorem of black  hole is declared to be ``a stationary asymptotically flat axisymmetric solution of Einstein gravity must be Kerr." In this sense, we find a counterexample of this claim. Through  detailed analysis of the theorem we find that our result in fact does not violate the Israel and Carter theorems. To find the observational indications is the future work. This Letter implicitly displays that the Einstein gravity may still hide more amazing structures beyond our present studies.

\section{Acknowledgements}
This work is supported by the Program for Professor of Special Appointment (Eastern Scholar) at Shanghai Institutions of Higher Learning, National Education Foundation of China under grant No. 200931271104, and National Natural Science Foundation of China under Grant No. 11075106 and 11275128.

\end{document}